\documentclass[aps,pre,amsmath,amssymb]{revtex4-1}
\usepackage{graphicx}

\usepackage{epsf}

\newcommand{\be}{\begin{equation}}
\newcommand{\ee}{\end{equation}}
\newcommand{\ba}{\begin{array}{l}}
\newcommand{\ea}{\end{array}}
\newcommand{\banonum}{\begin{eqnarray*}}
\newcommand{\eanonum}{\end{eqnarray*}}
\newcommand{\baa}{\begin{eqnarray}}
\newcommand{\eaa}{\end{eqnarray}}
\newcommand{\ed}{\end{document}}
\newcommand{\lab}[1]{\label{#1}}
\newcommand{\re}[1]{(\ref{#1})}
\newcommand{\ci}[1]{\cite{#1}}
\newcommand{\bfr}{\begin{flushright}}
\newcommand{\efr}{\end{flushright}}
\newcommand{\bfl}{\begin{flushleft}}
\newcommand{\efl}{\end{flushleft}}

\renewcommand{\baselinestretch}{1.4}
\date{}
\begin{document}

\title{Finite-temperature nonlinear dynamics in cavity QED:
\\A Thermofield Dynamics Approach}
\author{D.U.Matrasulov\thanks{E-mail: dmatrasu@phys.ualberta.ca;},
T.Ruzmetov, D.M.Otajanov, P.K.Khabibullaev, A.A.Saidov\\
Heat Physics Department of the Uzbek Academy of Sciences,
\\ 28 Katartal St.,700135 Tashkent, Uzbekistan\\
and\\
F.C.Khanna\thanks{E-mail: khanna@phys.ualberta.ca;}\\
Physics Department University of Alberta\\
Edmonton Alberta, T6G 2J1 Canada\\}

\begin{abstract}
Heath-bath effects in the dynamics of  atom + cavity system are
studied. The temperature effects are explored using thermofield
dynamics formalism. It is found that the dynamics of the system is
sensitive to small changes in the temperature and the thermal
effects lead to increasing instabilities by causing transitions
from regular to chaotic motion.
\end{abstract}
\maketitle

\section{Introduction}
Cavity quantum electrodynamics deals with studying the interaction
of atoms with photons in high-finesse cavities in a wide range of
the electromagnetic spectrum, from microwaves to visible light.
The fact that the system "atom + cavity mode" is a quantum system
makes cavity quantum electrodynamics (QED) an excellent testing
ground for such important issues of modern quantum physics as
quantum measurement theory, entanglement, quantum computation,
quantum interference and at the same time provides a unique
possibility for trapping, cooling and manipulating of atoms
\ci{spec}.. Practical importance of cavity QED is mainly related
to potential possibility for manipulating atoms and photons in
mesoscopic scales. Therefore, in recent years cavity QED has
become one of the hot topics both in theoretical and experimental
context \ci{qed,spec,adams}. In particular, a number of new
phenomena, such as the realization of quantum phase gate
\ci{turch}, creation of Fock states of the radiation field
\ci{muns,kim} and quantum non-demolition measurements \ci{spec}
have been found. The dynamics of a single atom trapped in a
microcavity is governed by quantum electrodynamics. This makes
cavity QED an interdisciplinary area as many subfield of physics,
such as quantum and atomic optics, cold atom physics, physics of
nanosized systems and quantum information, may use important
results of the cavity QED.

Recently cavity QED is considered in dealing with nonlinear
dynamics \ci{prants1,zas}. Mapping quantum equations of motion
onto classical ones, for the Jaynes-Cummings Hamiltonian, which
includes recoil motion of the atom, Prants. et. al, explored
phase-space dynamics of the atom interacting with a single cavity
mode by analyzing Poincare surface sections and calculating
Lyapunov exponents. In such a case classical and not quantum chaos
described by classical equations of motion corresponding to
Jaynes-Cummings hamiltonian are studied that are classical
counterpart of the atom+photon+cavity system that are described by
Prants et al \ci{prants1}-\ci{prants3}. This system can be
considered as a periodically driven quantum system as most of the
experiments on the study of chaos in atoms explore interaction of
cold atoms with the standing waves of various type, such as
phase-modulated standing wave \ci{moor}, amplitude-modulated
standing wave and pulsed standing wave \ci{hens}.

In this paper we extend the approach developed by Prants et al.
\ci{prants1,zas} to the case of the system coupled to a heat bath
i.e. at finite temperature. For this purpose we use a real-time
finite-temperature quantum field theory, thermofield dynamics
\ci{um,tak,das}. In particular, by solving temperature-dependent
classical equations of motion we explore Poincare surfaces
sections, Lyapunov exponents and Levy flights in
finite-temperature cavity QED. The role of a thermal bath, noise
and dissipation effects in cavity QED is of importance both from
fundamental and practical viewpoints. This paper is organized as
follows: in the next section we present brief description of TFD
formalism. Section 3 briefly recalls the results for $T=0$
obtained by Prants et al. Section 4 presents treatment of "atom +
confining mode" system at finite temperature, while numerical
results for Poincare surfaces of section, Levy flights and
Lyapunov exponent at $T=0$ are given in section 5. Finally,
section 6 presents some concluding remarks.

\section{Thermofield Dynamics}

TFD is a real time operator formalism of quantum field theory at
finite temperature in which any physical system can be constructed
from a temperature-dependent vacuum which is a pure state
\ci{um}-\ci{ademir}. For finite temperature it has been recognized
for a long time that the Hilbert space has to be doubled. This is
achieved with Thermofield Dynamics. One of the objective was to
get a theory that real time and at the same time it is at finite
temperature. The second Hilbert space is introduced along with a
set of operators that are similar to be distinct from the normal
operators. The final results don't depend explicitly on this
second set of operators. However, Bogoliubov transformation mixes
the two set of operators.This brings in factors that are
temperature-dependent.The vacuum is also temperature-dependent and
the creation and annihilation operators are converted to
temperature-dependent operators, such that the annihilation
operator acting on this vacuum gives zero.
  For a long time it was thought that these operators are
ghosts. However, Umezawa showed that these operators can be viewed
as degrees of freedom of the heat bath in the classical theory
\ci{um}. Later on it was shown by Santana and Khanna that one set
gives the set of observables and the other gives symmetry
operators. This argument is based on group theory and is given in
the ref. \ci{ademir}

Thus TFD is a powerful tool for exploring quantum dynamics of a
system at finite temperature provided its Hamiltonian can be
represented in terms of annihilation and creation operators. It
has found many applications in condensed matter physics
\ci{um,tadic,egor}, especially in superconductivity theory and
related topics. Recently TFD has been applied to explore quantum
chaos in the Yang-Mills-Higgs system \ci{matr} and for the
calculation of the spectra of a strongly interacting bound system
\ci{matr1}. In this work we apply TFD prescription to
Jaynes-Cummings model. It should be noted that TFD has been
applied earlier  to Jaynes-Cummings model \ci{barnet} where the
thermal noise effects in quantum optics are studied.

 Applying TFD prescription to a quantum system implies
performing two actions: \ci{tak,das} \\
i) doubling of the Fock space which means that all operators are
doubled by introducing their tilded partners
describing the effects of a heath bath; and\\
ii) using Bogoliubov transformations, which makes Hamiltonian of
the system temperature-dependent . With respect to a Hamiltonian
operator written in terms of annihilation and creation operators,
doubling means that the total Hamiltonian is written as \be \hat H
= H - \tilde H, \ee while the Bogoliubov transformations which are
given by
$$
 a=a(\beta)\cosh\theta + \tilde a^{ \dagger}
(\beta)\sinh\theta
$$
$$
 a^{\dagger} =a^{\dagger}(\beta)\cosh\theta + \tilde a
(\beta)\sinh\theta
$$
\be
 \tilde a = a^{ \dagger}(\beta)\sinh\theta + \tilde a
(\beta)\cosh\theta \ee
$$
\tilde a^{\dagger} = a(\beta)\sinh\theta + \tilde a^{\dagger}
(\beta)\cosh\theta,
$$
 make this Hamiltonian temperature-dependent.

Here \be \beta = \frac{\omega}{k_B T}; \;\;\;\; \sinh^2 \theta =
(e^\beta - 1)^{-1} \ee
 and the annihilation and creation operators
satisfy the following commutation  relations:

\be
 [a(\beta),
a^+(\beta)] =1\;\;\;\; [\tilde a(\beta), \tilde a^+(\beta)] =1 \ee

 All other commutation relations are zero.

We note that we are dealing with a system in equilibrium, i.e. the
temperature, $T$, is constant. Since the Hamiltonian of the
Jaynes-Cummings model is written in terms of annihilation and
creation operators, it is convenient to use TFD formalism to treat
this model at finite temperature. Applying the above prescription
to the Jaynes-Cummings model we transform the Hamiltonian of the
system into the temperature-dependent form that allows us to treat
chaos using the same approach as that used by Prants et.al
\ci{zas}.

\section{Cavity QED at $T=0$}

In this section we briefly recall the case of $T=0$ which
is recently explored in detail in a series of papers by Prants et.al. \ci{prants1}-\ci{prants5}.
The simplest cavity QED system is a single two-level atom
interacting with a single standing wave mode moving along the
$x$-axis with the frequency $\omega_f$.

The dynamics of this system in the presence of atomic recoil
motion is described by Jaynes-Cummings Hamiltonian, which is
written as
\be \hat{H}=\frac{\hat{P^2}}{2m}+\hbar\omega _a
\hat{S_z}+\hbar\omega_f\hat{a^+}\hat{a}-\hbar\Omega_0
(\hat{a^+}\hat{S_-}+\hat{a}\hat{S_+})\cos(k_f\hat{x}),
\lab{hamilt} \ee

where $\hat S_z $, $\hat S_+ $ and $\hat S_- $ - are expressed in
terms of pauli matrices i.e. $\hat S_{+,-} = ( \hat S_x , \hat S_y
)$ and $\hat S_z $ is the $z$ - component, and $\hat{a^+}$ and
$\hat{a}$ are the creation and annihilation operators,
respectively, describing a selected mode of the radiation field of
the frequency $\omega_f$ and the wave number $k_f$ in a lossless
cavity. The parameter $\Omega_0$ is the amplitude value of the
atom-field dipole coupling and depends on the position of an atom
inside a cavity. To treat  the nonlinear dynamics of this system
Prants et. al. derived first the quantum equations of motion for
the operators external atomic operators, $P$ and $x$ and slowly
varying amplitudes of the field and spin operators: $\hat a(t)
=\hat a e^{-i\omega_ft},$ $\hat a^+(t) =\hat a^+ e^{i\omega_ft}$,
$\hat S_{\pm}(t) =\hat S_{\pm} e^{\mp i\omega t}$ and $\hat
S_{z}(t) =\hat S_{z}$.

Taking the averages of all operators over an initial quantum
state(which is  a product of the translational, electronic, and
the radiation field states) the quantum equations of motions can
be replaced by the equations for the expectation values of the
operators \ci{zas}. The equations of motions for these averages
are written as
$$
\dot{x}= \alpha p
$$
$$
\dot{p}=-2(a_{x}s_{x}+a_{y}s_{y})sin{x}
$$
$$
\dot{s}_{x}=-\delta s_{y}+2a_{y}s_{z}cos{x}
$$
\be
\dot{s}_{y}=\delta s_{x}-2a_{x}s_{z}\cos{x}
\lab{eqmot}
\ee
$$
\dot{a}_{x}=-s_{y}\cos{x}
$$
$$
\dot{a}_{y}=s_{x}\cos{x},
$$
where the expectation values are defined as
$$
 x=k_f\langle\hat x
\rangle,p=\langle\hat p\rangle/\hbar{k}_f
$$
$$
 s_x= \langle s_-+s_+\rangle/2,s_y=\langle s_- -s_+\rangle/2i
$$
\be
 a_x=\langle a+a^+\rangle/2,a_y=\langle a-a^+\rangle/2i,
\lab{exp} \ee
$$
 \alpha=\hbar k_f^2 /m\Omega_o, \tau=\Omega_o t
$$

In the ref. \ci{zas} the dynamics of the cavity+atom system at
$T=0$ is treated by solving Eq. \re{eqmot}  and analyzing the
solutions in terms of Poincare surface sections and Lyapunov
exponents. The extension of the zero temperature results \ci{zas}
to non-zero temperature is done here by using the formalism of
TFD.

\section{Cavity QED at non-zero temperature}

Applying TFD prescription to the cavity QED Hamiltonian
\re{hamilt}, we have

$$
\hat{H} = \frac{P^2-\tilde{P^2}}{2m} +
 \hbar \omega_f ( a^+(\beta) a(\beta) - \tilde{a}^+(\beta) \tilde{a} (\beta))
$$
$$
+ \hbar\Omega_0 [  (a(\beta)\sinh\theta -
\tilde{a} (\beta)\sinh\theta +
 \tilde{a}^+ (\beta)\cosh\theta -
a^+(\beta)\cosh\theta) S_-
$$
\be
 + (a^+(\beta)\sinh\theta - \tilde{a}^+(\beta)\sinh\theta
\tilde{a}(\beta) + \cosh\theta - a(\beta)\cosh\theta)S_+ ]
\cos(k_f\hat{x}).
\ee

Repeating the same steps as those used in \ci{zas} for $T=0$ we
have the temperature-dependent equations of motion for the
expectation values of coordinate, momentum, spin, annihilation and
creation operators
$$
\frac{d\hat{x}}{dt}=\frac{P-\tilde{P}}{m}
$$
$$
\frac{d \hat{P}}{dt}= \hbar k_{f}
\Omega_{0}[ ( a^+(\beta)\cosh\theta - \tilde{a}^+(\beta)\cosh\theta-
a(\beta)\sinh\theta + \tilde{a}(\beta)\sinh\theta)S_-
$$
$$
+(a(\beta)\cosh\theta+\tilde{a}^+(\beta)\sinh\theta-a^+(\beta)\sinh\theta-
\tilde{a}(\beta)\cosh\theta)S_+]\sin{k_f\hat{x}}
$$
$$
\frac{dS_+}{dt}=i(\omega_f-\omega_a )S_+
+2i\Omega_{0}S_z[a(\beta)\sinh\theta-\tilde{a}(\beta)\sinh\theta+
\tilde{a}^+(\beta)\cosh\theta-\tilde{a}(\beta)\cosh\theta]\cos{k_f\hat{x}}
$$
$$
\frac{dS_-}{dt}=-i(\omega_f-\omega_a ) S_ -
-2i\Omega_0 S_z[a^+(\beta)\sinh\theta-\tilde{a}^+(\beta)\sinh\theta+
\tilde{a}(\beta)\cosh\theta-a(\beta)\cosh\theta ]\cos{k_f\hat{x}}]
$$
$$
\frac{da^+}{dt}=-i\Omega_{0}(S_-\sinh\theta-S_+\cosh\theta)\cos{k_f\hat{x}}
$$
\be
\frac{da}{dt}=-i\Omega_0(S_-\cosh\theta-S_+\sinh\theta)\cos{k_f\hat{x}}
\lab{eq9}
\ee
$$
\frac{d\tilde{a}}{dt} =
-i\Omega_0(S_{+}\sinh\theta-S_{-}\cosh\theta)\cos{k_f\hat{x}}
$$
$$
\frac{\tilde{a^+}}{dt} =
i\Omega_0(S_{+}\cosh\theta-S_-\sinh\theta)\cos{k_{f}\hat{x}}
$$
$$
\frac{dS_z}{dt}=i\Omega_{0}(B_{1}S_{-}-B_{2}S_{+})\cos{k_f\hat{x}},
$$
 where
$$
B_1=a(\beta)\sinh\theta-\tilde{a}(\beta)\sinh\theta+\tilde{a}^+
(\beta)\cosh\theta-a^+(\beta)\cosh\theta,
$$
\be
B_{2}=a^+(\beta)\sinh\theta-\tilde{a}^+(\beta)\sinh\theta+
\tilde{a}(\beta)\cosh\theta-a(\beta)\cosh\theta.
\lab{eq10}
\ee

These equations  describe the time evolution of the dynamical
variables(expectation values of the operators) in the presence of
coupling to  a thermal bath with constant temperature, $T$. We
note that interaction of the spin degrees of freedom with a
thermal bath is not taken into consideration in these equations.

\section{Results and discussion}

We have solved numerically the system of equations \re{eq9} and
plotted Poincare surfaces of section (PSS) at various values of
temperature. In Fig. 1 the Poincare surface section are plotted
for: $\beta = 2 $ ($a$); $\beta = 6$ ($b$); $\beta = 10$ ($c$);
$\beta = 12$ ($d$).

It is clear from these plots that for high temperatures the
dynamics is fully chaotic, while by decreasing $T$ transition to
mixed and regular regime of motion can be observed. To compare the
approach for $T=0$ and our approach for finite $T$ in Fig. 2, PSS
are plotted by solving the set of equations \re{eq9} and \re{eq10}
($T \approx 0$). For $T$ about zero our results for PSS agree with
those by Prants \ci{zas}.

Following the prescription developed in \ci{zas} for exploring of
instabilities in the QED cavity, another characteristics of
chaoticity, so-called Levy flights \ci{hens,shles93,zas98} is
analyzed. As is well known, chaotic motion in classical system
have several quantitative and qualitative characteristics, such as
phase-space trajectories, Lyapunov exponent and Levy flights
\ci{shles93,zas98}. The latter are the pieces that appear in the
trajectory of a particle in a transition from a regular to a
chaotic regime of motion. In other words, Levy flights are the
chaotic pieces interrupting regular behaviour of the trajectory of
an oscillating or regularly moving particle \ci{zas,shles93}. In
Fig. 3, Levy flights for various values of $\beta$ are plotted. It
is clear from this figure that for (a) $T=0$  and (b) $\beta =
100$ the plots are similar, with the same number of flights.
However, by increasing the temperature ($\beta = 10$ and
$\beta=5$) leads to increasing number of flights.

To make our treatment more comprehensive
we should consider also the behaviour of the
maximum Lyapunov exponent at different temperatures. The maximum
Lyapunov exponent characterizes the mean rate of the exponential
divergence of initially close trajectories and serves as a
quantitative degree of deterministic chaos in the system. In Fig. 4
the maximum Lyapunov exponent
is plotted as a function of detuning parameter, $\delta$ at different values of temperature.
Again, one can observe
"more chaos" in the case of finite temperature compared to
$\beta=100$ case. For higher temperatures (smaller $\beta$) the
Lyapunov exponent becomes more higher than that for lower
temperatures.

In Fig. 5 the maximum Lyapunov exponent versus atom field detuning $\delta$
and initial atomic momentum $p_0$ is plotted for $T=0$ and $\beta
=0.5$. This plot also shows that increasing of the heat bath
temperature leads to increasing of maximum Lyapunov exponent for
all the values of $\delta$ and $p_0$. However, some "islands" near
$=0$ still exist in this plot. This means that Lyapunov exponent
remains as small near $\delta =0$ at $T \ne 0$.

In all cases(including the case of $T\ne 0$) one can observe that $\lambda$ becomes equal to zero at $\delta=0$ that means becoming
of our system integrable for  $\delta=0$.

Finally, Fig.6 presents maximum Lyapunov exponent versus $\delta$
and $\beta$. Again increasing of $\lambda$ for higher $T$ (smaller
$\beta$) can be observed.
 Therefore besides the control parameters $\alpha$ and
$\delta$, in the case of finite temperature we have an additional
parameter for controlling the dynamics of the atom in a cavity
temperature, $T$.

\section{Conclusion}

Thus we have studied finite-temperature nonlinear dynamics of an
atom coupled to a single mode of the cavity field. Applying the
formalism of a real-time finite-temperature field theory to the
Jaynes-Cummings Hamiltonian and using the same approach as that
used in \ci{zas} we have studied classical dynamics of the
"atom+cavity mode" system  coupling to a thermal
bath. The equations of motions for the classical dynamics are
obtained by "mapping" of quantum dynamics onto classical one as in the ref. \ci{zas}.

Using the temperature-dependent equations of motion,
dependance of the dynamics on heat-bath effects or finite
temperature effects are considered. The results show that the
dynamics is quite sensitive to the small changes of temperature.
Qualitative characteristics  of chaoticity of the system are
explored for different values of $\beta$. In particular,
projections of the Poincare surface section are plotted for
different values of temperature. It can be seen by comparing the
sensitivity of Poincare surface sections, Levy flights and the
maximum Lyapunov exponent to the changes of $\beta$ with those of
$\delta$ that the dynamics is more sensitive to the changes of
temperature than that of $\delta$. This implies that the
temperature of a thermal bath can be considered as an additional
control parameter for the dynamics of an atom coupling to a cavity
mode.

\section{Acknowledgments}
This work is supported by the INTAS YS Fellowship (Ref. Nr.06-1000023-6008 ) and by the grant of Volkswagen Foundation (Ref Nr. I/82 136).
The work of DMO is supported by the grant of the uzbek Academy of Sciences (FA-F2-084).
The work of FCK is supported by NSERCC.

\newpage

\begin{figure}[htbp]
\begin{center}

\includegraphics[width=14cm,height=14cm]{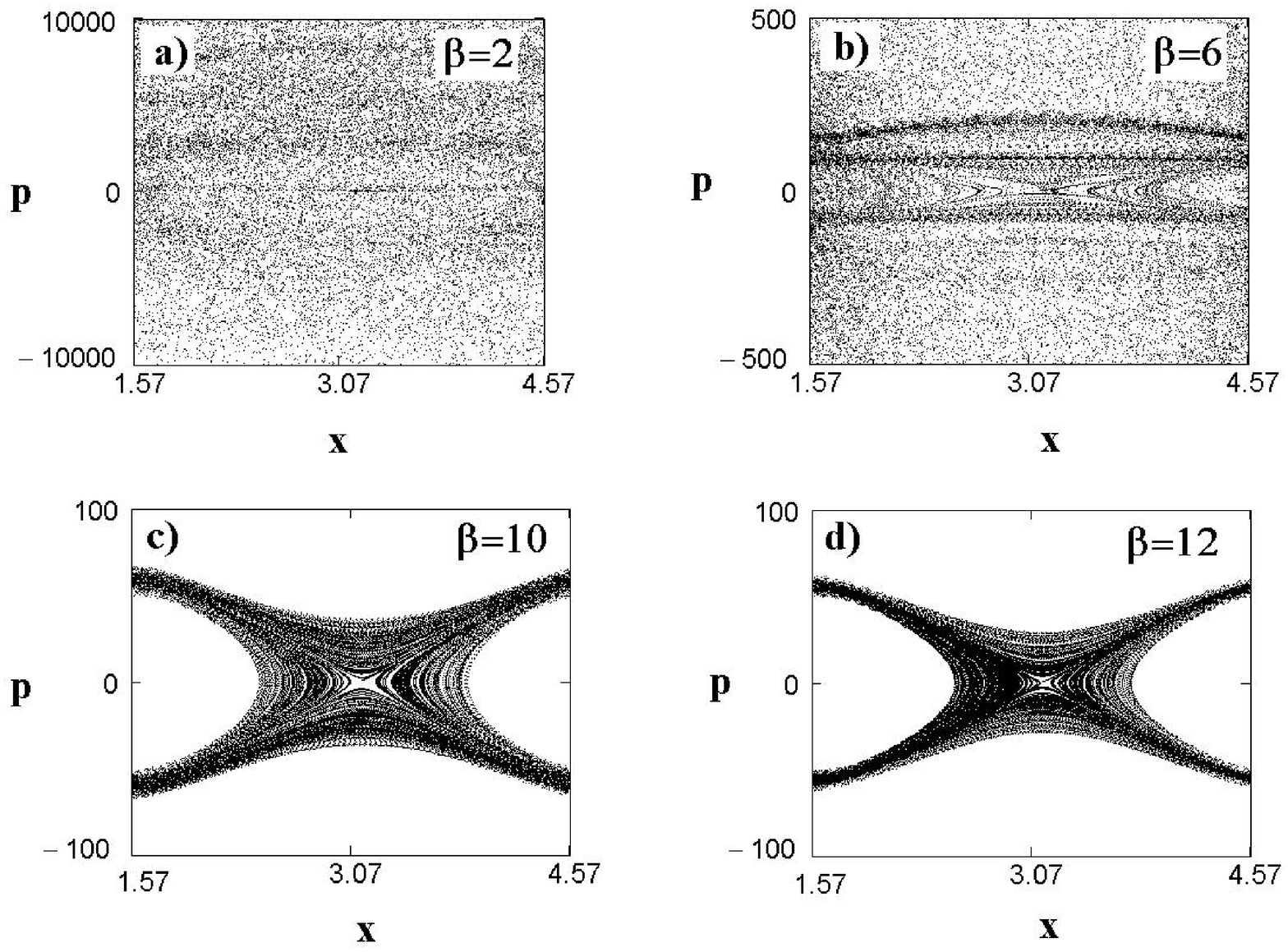}
\vspace{.5cm} \caption{Projection of the Poincare sections at
finite temperature on the plane of the atomic momentum $p$ in
units $\hbar k_f$ and the position in units of $k_f^{-1}$.  (a)
$\beta = 2$ ; (b) $\beta = 6$; (c) $\beta = 10$; (d) $\beta = 12$.
$\delta=1.92$ and $s_z(0)=-0.863$ in all cases. $x,p$ are
dimensionless.}
\end{center}
\end{figure}

\newpage

\begin{figure}[htbp]
\begin{center}

\includegraphics[width=14cm,height=7cm]{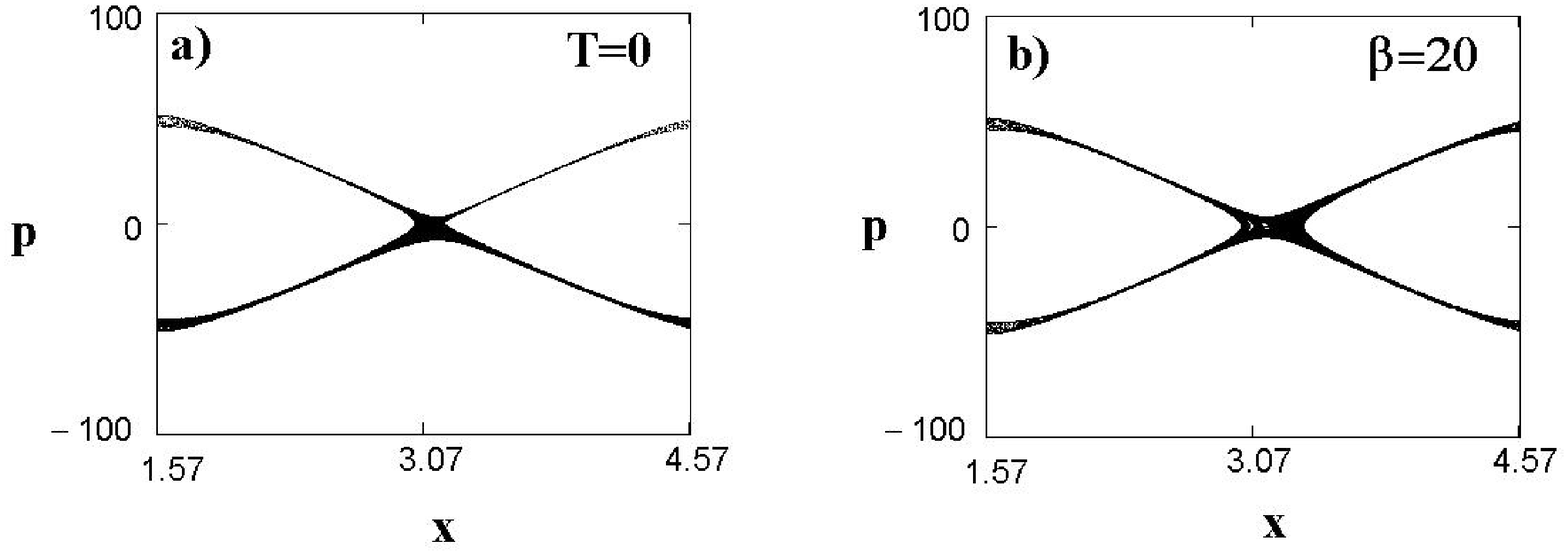}
\vspace{.5cm} \caption{Projection of the Poincare sections at
finite (b) and zero (a) temperature on the plane of the atomic
momentum $p$ in units $\hbar k_f$ and the position in units of
$k_f^{-1}$. (a) $T = 0$; (b) $\beta = 20$. $\delta=1.92$ and
$s_z(0)=-0.8660254$ in both cases. $x,p$ are dimensionless.}
\end{center}
\end{figure}

\newpage

\begin{figure}[htbp]
\begin{center}

\includegraphics[width=14cm,height=12cm]{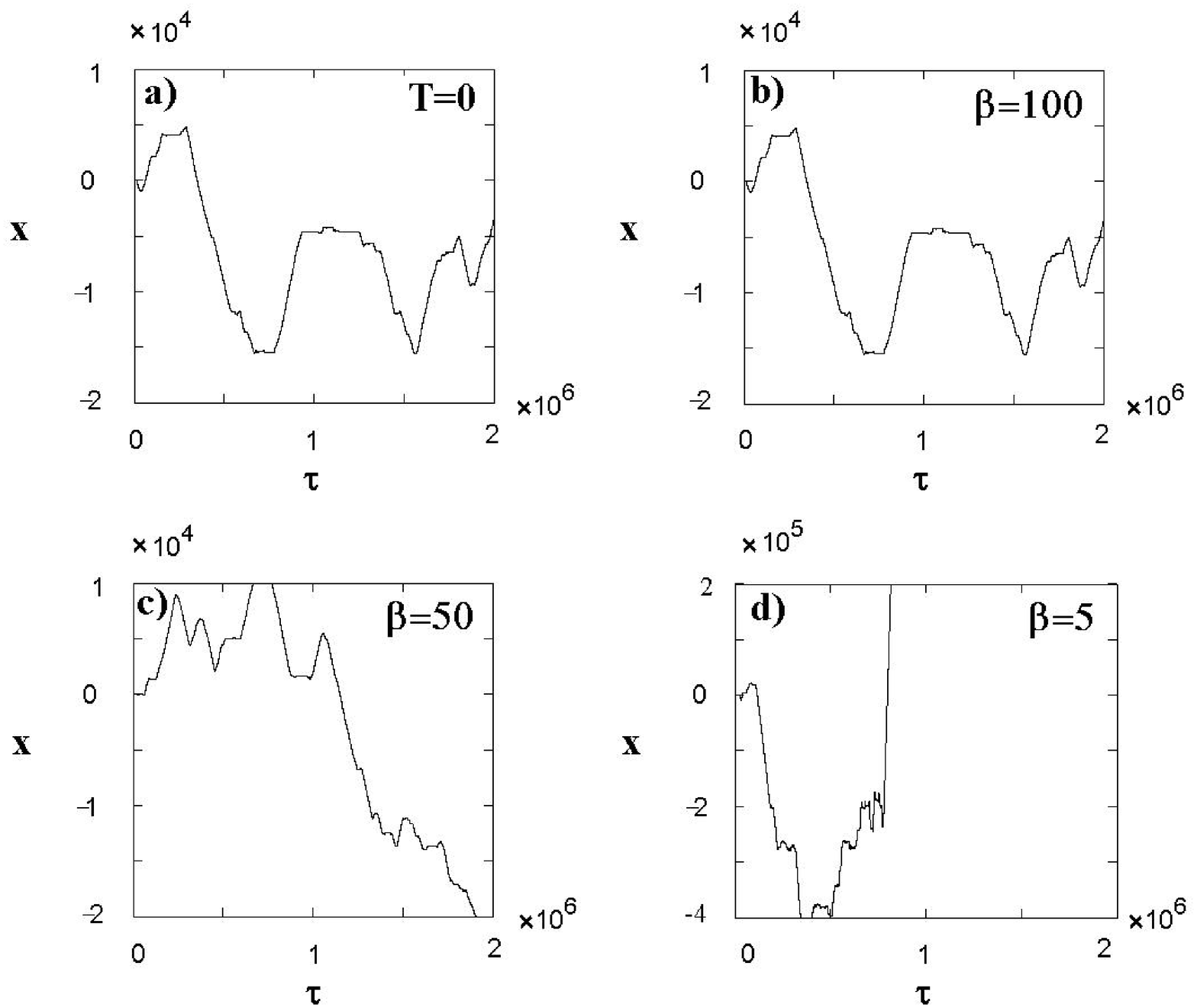}
\vspace{.5cm} \caption{  Levy flights of an atom in a cavity at
zero (a) and finite (b,c,d) temperatures. a) $T = 0$; b) $\beta =
100$; c) $\beta = 50$; d) $\beta = 5$. $\delta=1.2$ and
$s_z(0)=-0.8660254$ in all cases. Time is in units of
$\Omega_0^{-1}$. $x,p,\tau$ are dimensionless.}

\end{center}
\end{figure}

\newpage

\begin{figure}[htbp]
\begin{center}
\includegraphics[width=15cm,height=10cm]{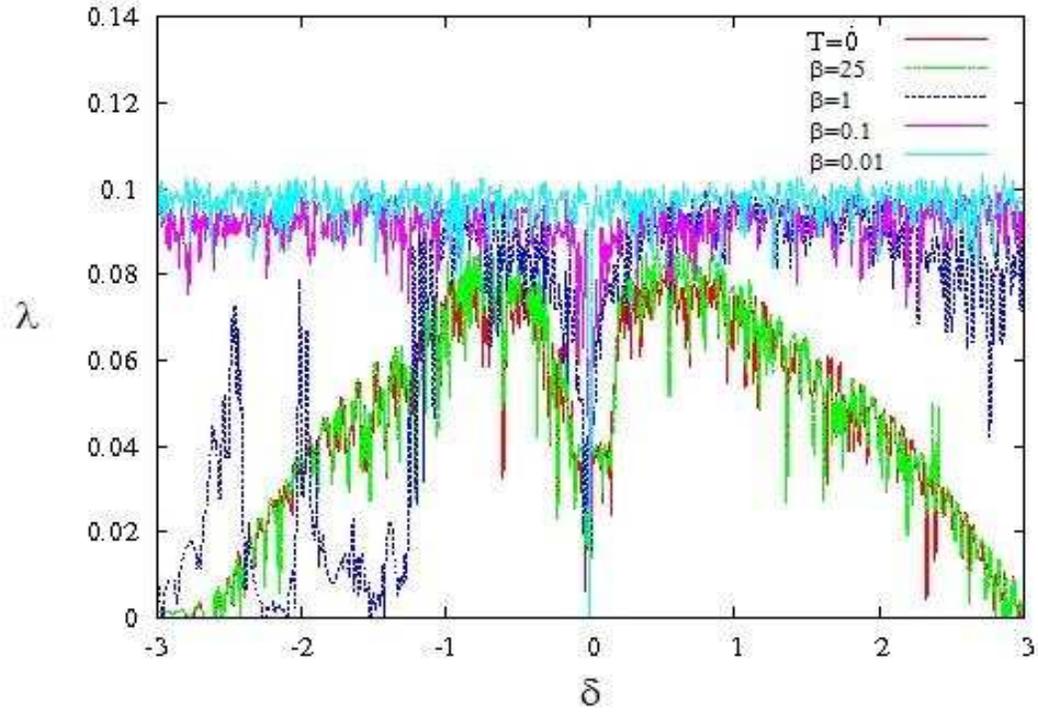}
\vspace{.5cm} \caption{(Color online) The maximum Lyapunov exponent $\lambda$ in
units of maximal atom-field coupling rate $\Omega_0$ versus the
atom-field detuning $\delta$ in units $\Omega_0$ at zero and
finite temperatures:
 $\beta = 25$; $\beta = 1$;  $\beta = 0.1$; $\beta=0.01$
and $s_z(0)=0$ in all cases.}
\end{center}
\end{figure}

\newpage

\begin{figure}[htbp]
\begin{center}
\includegraphics[width=12cm,height=8cm]{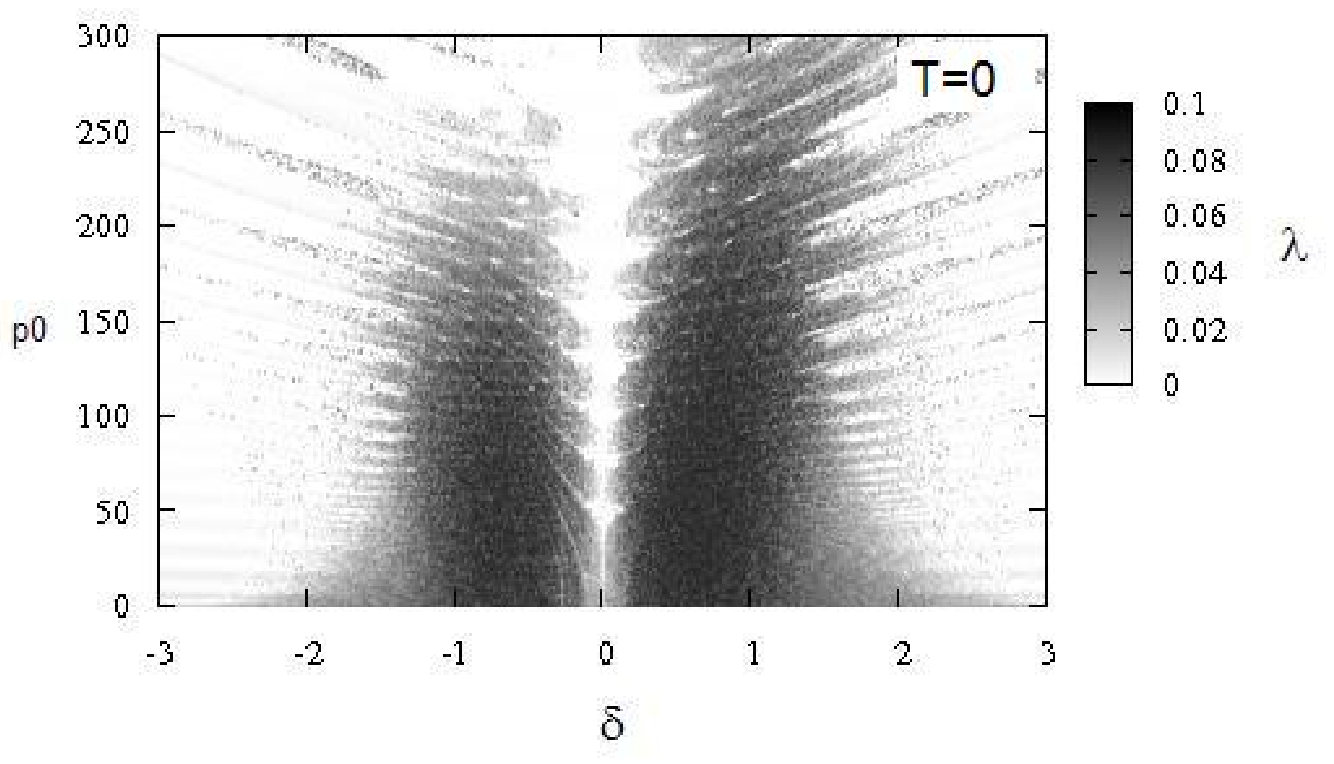}
\includegraphics[width=12cm,height=8cm]{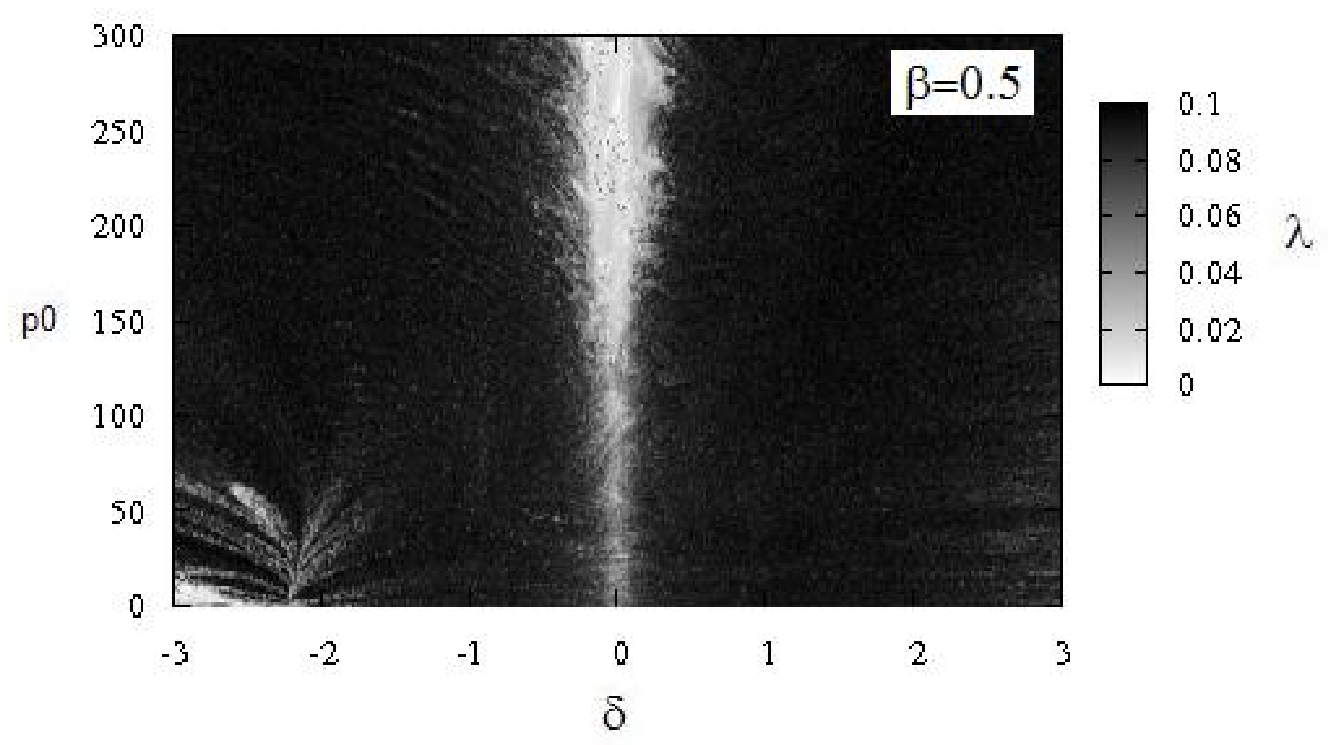}
\vspace{.5cm} \caption{The maximum Lyapunov exponent $\lambda$ (in
units of maximal atom-field coupling rate $\Omega_0$) versus the
atom-field detuning $\delta$ (in units $\Omega_0$) at different
temperatures: $T=0$; and $\beta =0.5$; ($s_z(0)=0$ in
both cases).}
\end{center}
\end{figure}

\newpage

\begin{figure}[htbp]
\begin{center}
\includegraphics[width=12cm,height=8cm]{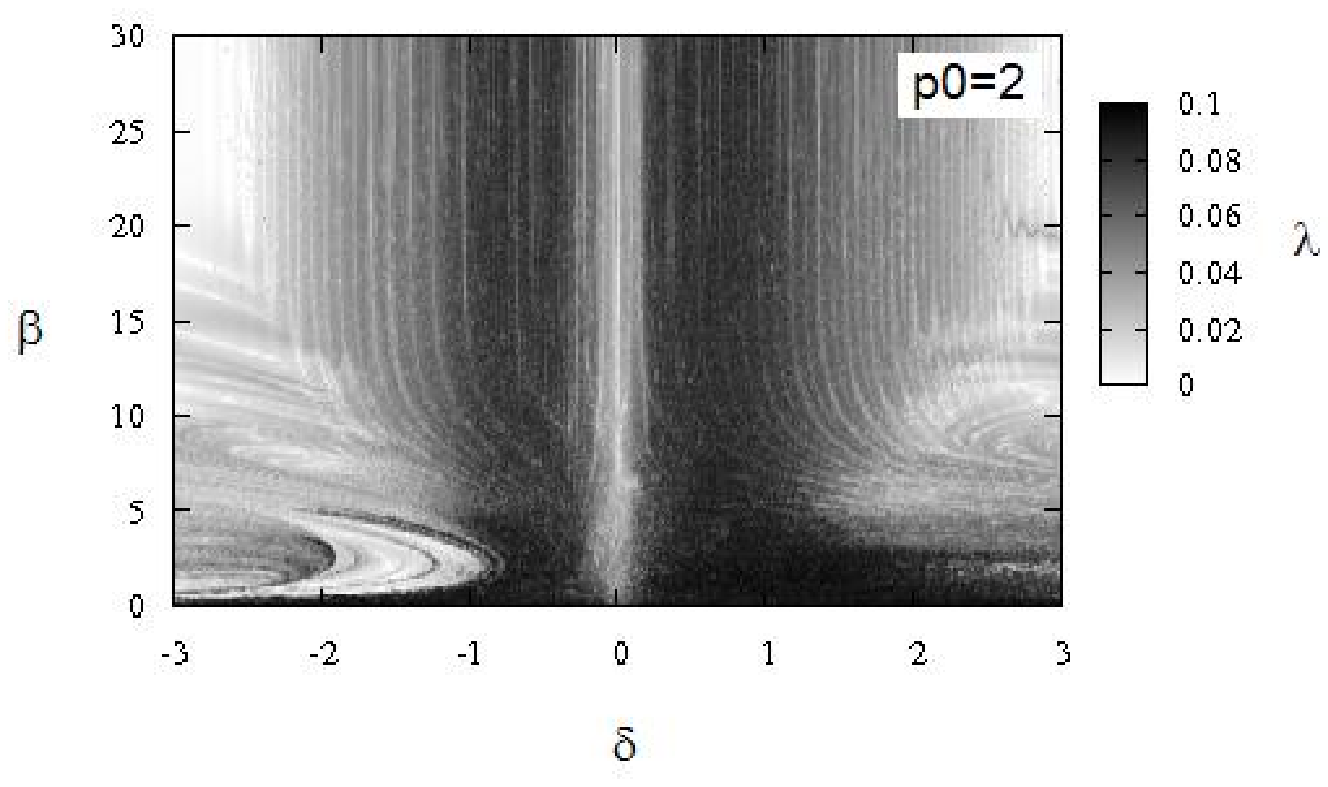}
\vspace{.5cm} \caption{The maximum Lyapunov exponent $\lambda$ (in
units of maximal atom-field coupling rate $\Omega_0$) versus the
atom-field detuning $\delta$ (in units $\Omega_0$) and $\beta$;
($p_0 =2$ and $s_z(0)=0$).}
\end{center}
\end{figure}


\begin{thebibliography}{99}
\bibitem{qed} \emph{Cavity Quantum Electrodynamics. Edited by P.R.Berman} (Academic, New York 1994)
\bibitem{spec} Special Issue on Modern Studies of Basic Quantum Concepts \emph{Phys. Scr.} {\bf T76} (1998)
\bibitem{adams} C.S.Adams, M.Sigel, J.Mlnek \emph{Phys. Rep.} {\bf 240} 143 (1994)
\bibitem{turch} Q.A.Turchette \emph{et. al. Phys. Rev. Lett.} {\bf 75} 4710(1995)
\bibitem{muns} P.Munsterman,T.Fischer,P.Maunz,P.W.H.Pinkse,and G.Rempe \\
 \emph{Phys.Rev.} {\bf 82} {1999}
\bibitem{kim} R.Miller, T.E.Northup,K.M.Birnbaum,A.Boca,A.D.Boozer and H.J.Kimble\\
\emph{J.Phys B} {\bf 38}(2005)S551-565S
\bibitem{prants1} S.V. Prants and L.E. Kon'kov \emph{Chaos, Solitons \& Fractals } {\bf 11} 871 (2000)
\bibitem{prants2} V.I.Ioussoupov, L.E. Konkov and S.V. Prants \emph{Physica D} {\bf 155} 311 (2001)
\bibitem{zas} S.V.Prants, M.Edelman, G.M.Zaslavsky,  \emph{Phys. Rev. E} {\bf 66} 046222 (2002)
\bibitem{prants3} S.V. Prants and M. Yu. Uleysky \emph{Phys. Lett. A}, {\bf 309} 357 (2003)
\bibitem{prants4} S.V. Prants,  M.Yu. Uleysky and Yu. Argonov, \emph{Phys. Rev. A.} {\bf 73} 023807  (2006)
\bibitem{prants5} V.Yu. Argonov and S.V. Prants,   Phys. Rev. A. {\bf  75} 063428   (2007)
\bibitem{um}  H. Umezawa,  H. Matsumoto and M. Tachiki \emph{Thermofield Dynamics.}(North-Holland. Amsterdam, 1982)
\bibitem{tak} Y. Takahashi and H. Umezawa,  \emph{Int.J. Mod.Phys.B} {\bf  10}  1755 (1996).
\bibitem{das} Ashok Das, \emph{Finite Temperature Field Theory}. (World Scientific, New York, 1977)
\bibitem{tadic} B. Tadic, R. Pirc, R. Blinc, \emph{Physica B}, {\bf 168}  85 (1990)
\bibitem{egor} B.V. Egorov, \emph{J. Phys.: Condens. Matter} {\bf 4} 4115 (1992)
\bibitem{matr} D.U. Matrasulov, F.C. Khanna, U.R. Salomov and A.E. Santana \\
\emph{Eur. Phys. J. C} {\bf 42} 148 (2005)
\bibitem{matr1} D.U. Matrasulov, F.C. Khanna, Kh.T. Butanov and Kh.Yu. rakhimov \\
 \emph{Mod.Phys.Lett. A} {\bf 21} 1383 (2006)
\bibitem{ademir} A.E.Santana, F.C.Khanna, \emph{Phys. Lett. A} {\bf 203} 68 (1995)
\bibitem{barnet} S.M.Barnett and P.L.Knight \emph{J. Opt. Soc. Am. B} {\bf 2} 467 (1985)
\bibitem{moor}  F.L. Moore, J.C. Robinson, C. Bharucha, P.E. Williams, and M.G. Raizen, Phys. Rev. Lett. {\bf 73} 2974 (1994)
\bibitem{hens} W.K. Hensinger, A.G. Truscott, B. Upcroft, M. Hug, H.M. Wiseman, N.R. Heckenberg, and H. Rubinsztein-Dunlop,
Phys. Rev. A {\bf 64} 033407 (2001)
\bibitem{shles93} M.F.Shlesinger, G.M.Zaslavsky, and J.Klafter, \emph{Nature} {\bf 363} 31 (1993)
\bibitem{zas98} G.M.Zaslavsky, \emph{Physics of Chaos of Hamiltonian Systems} (Imperial
College Press, London, 1998)

\end{thebibliography}
\end{document}